\documentclass[aps,draft]{revtex4}

\input epsf

\begin{document}

\title{
Fermionic Quasiparticle Representation of Tomonaga-Luttinger Hamiltonian
}
\author{A.V. Rozhkov}

\affiliation{
Institute for Theoretical and Applied Electrodynamics JIHT RAS,
Moscow, ul. Izhorskaya 13/19, 127412, Russian Federation
}

\begin{abstract}
We find a unitary operator which asymptotically diagonalizes the
Tomonaga-Luttinger hamiltonian of one-dimensional spinless electrons. The
operator performs a Bogoliubov
rotation in the space of electron-hole pairs. If bare interaction of the
physical electrons is sufficiently small this transformation
maps the original Tomonaga-Luttinger system on a system of free fermionic
quasiparticles. Our representation is useful when the electron dispersion
deviates from linear form. For such situation we obtain non-perturbative
results for the electron gas free energy and the density-density
propagator.
\end{abstract}

\maketitle
\hfill

\section{Introduction}
Bosonization is a standard approach to the problem of interacting 
electrons in one dimension (1D) \cite{haldane,boson}. Bosonization maps the
low-energy spectrum of the Tomonaga-Luttinger (TL) model on the spectrum of
free bosons.

In this paper we discuss a new method of treating such system.
We explicitly construct a unitary operator $U$ diagonalizing TL
hamiltonian.
Our unitary transformation maps the TL model on a system of free fermionic
quasiparticles. The description of the TL model in terms of these
quasiparticles
has several advantages over bosonization, as we will see below.

The diagonalizing procedure is closely related to the bosonization. The
operator $U$ can be thought of as a Bogoliubov transformation in the space
of particle-hole pair excitations. Alternatively, one can describe the action
of $U$ as a sequence of bosonization, Bogoliubov transformation which
diagonalizes the bosonic hamiltonian and further refermionization (fig.1).

When the dispersion of the physical electrons is linear both the 
quasiparticles
and the bosons offer equally good description of the TL model. If the
non-linear terms are substantial the free boson representation breaks down.
The fermionic quasiparticles show more resilience toward deviations from
the linear dispersion. They remain free as long as the bare interaction
constant is sufficiently small.

This latter property of the quasiparticle representation allows for
non-perturbative calculation of the free energy and the density-density
correlation function for the TL model with the non-linear dispersion. We
believe that these two results are new.

When the bare electron dispersion is linear the quasiparticle
representation could be used to determine the single-electron Green's
function. It coincides with the Green's function
obtained by other methods.

The existence of the quasiparticles does not contradict to the fact that
the
single-electron Green's function of TL model has no pole. The
quasiparticles
in TL model has zero overlap with the physical electrons: $\sqrt{Z} = 0$.
Therefore, it is convenient to think about TL model as $Z=0$ Fermi liquid
\cite{carmelo}.

The paper is organized as follows. We diagonalize the TL hamiltonian in
Sec. II. In Sec. III we offer an intuitive explanation to the method. In
this section all
technical complications are disregarded in order to create an integral
view. The density-density propagator is derived in Sec. IV. The
single-particle Green's function is calculated in Sec. V. Sec. VI is
reserved for discussion. Certain technical details can be found in Appendices.

\section{Diagonalization of TL hamiltonian}

The TL model hamiltonian is given by:
\begin{eqnarray}
H = H_{\rm kin} + H_{\rm int},\label{H}\\
H_{\rm kin} = i v_{\rm F}\int_{-L/2}^{L/2} dx 
\left( \colon\psi^\dagger_{{\rm L}}
\nabla\psi^{\vphantom{\dagger}}_{{\rm L}}\colon - \colon\psi^\dagger_{{\rm R}}
\nabla\psi^{\vphantom{\dagger}}_{{\rm R}}\colon \right) ,\label{H_kin}\\
H_{\rm int} = \int dx dx' \hat g(x-x') \rho_{\rm L}(x) \rho_{\rm R}(x'),
\label{int}\\
\rho_{\rm L,R} = \colon\psi_{\rm L,R}^\dagger \psi_{\rm
L,R}^{\vphantom{\dagger}}\colon.
\end{eqnarray}
The chirality label `L' corresponds to left-moving electrons, the label `R'
corresponds to right-moving electrons. The interaction of the electrons of
the same chirality is ignored since up to irrelevant operators such
interaction simply renormalizes the value of the Fermi velocity $v_{\rm F}$.
The symbol $:\ldots :$ denotes normal ordering of the field operators
$\psi$. A brief discussion of the normal ordering procedure we use in this
paper is given in Appendix A.

It is assumed that the cut-off of (\ref{H}) is infinite. To remove
ultraviolet divergences of the theory without cut-off we
replace usual zero-range interaction ($\hat g(x)=g_0\delta(x)$) by 
interaction acting over a finite range. Specifically, $\hat g(x)=g_0
\delta_\Lambda(x)$ where $\delta_\Lambda (x)$ is a broadened
version of the delta-function: its Fourier transform $\delta_\Lambda(q)$ is
such that $\delta_\Lambda(q)=1$ for $|q|$ smaller than some quantity
$\Lambda$, and $\delta_\Lambda (q)$ vanishes quickly for $|q|>\Lambda$.
The parameter $\Lambda$ thus defined regularizes ultraviolet divergences of
our theory.

It is easy to demonstrate (see, for example, \cite{haldane}) that the
following commutation relations are obeyed:
\begin{eqnarray}
\left[ \rho_{pq}, \rho_{p'-q'} \right] = \delta_{pp'} \delta_{qq'} p n_q
\label{rho_comm}\\
\left[\rho_{pq}^{\vphantom{\dagger}}, \psi^\dagger_{p'}(x) \right] =
\delta_{pp'} {\rm e}^{-iqx} \psi^\dagger_{p'}(x), \label{psi_comm}
\end{eqnarray}
where $p=+1$ for left-moving electrons, $p=-1$ for right-moving electrons
and
\begin{equation}
\rho_{pq} = \int dx {\rm e}^{-iqx} \rho_p (x),\quad n_q = \frac{Lq}{2\pi}.
\end{equation}
Using this commutation relations we will show that the unitary operator
\begin{eqnarray}
U={\rm e}^{\Omega},\\
{\Omega} = \sum_{q \ne 0 } \sum_p \alpha_q
\frac{p}{n_q} \rho_{p-q} \rho_{-pq}\label{Omega}
\end{eqnarray}
diagonalizes the hamiltonian (\ref{H}) except for the zero mode part.
Fortunately, the zero modes 
\begin{eqnarray}
N_{p} = \rho_{{p}q} \Big|_{q=0}
\end{eqnarray}
are decoupled from other degrees of freedom.
Also, their contribution to the low-energy spectrum is ${\cal O} (1/L)$.

Since $\rho$'s are quadratic in $\psi$
the above operator $U$ is quartic in fermionic operators $\psi$. In
general, it is impossible to work with such a non-linear
object. In our situation, however, the simplicity of commutation rules
(\ref{rho_comm})  and (\ref{psi_comm}) allows us to diagonalize the
hamiltonian with the help of $U$.

In order to transform the interaction term (\ref{int}) with the operator
$U$ it is enough to observe that the action of $U$ on the density operator
$\rho_{pq}$, $q \ne 0$, is a Bogoliubov rotation:
\begin{eqnarray}
U \rho_{pq} U^\dagger = u(\alpha_q) \rho_{pq} + v(\alpha_q) \rho_{-pq},
\label{BT}\\
u(\alpha_q) = u_q = \cosh(\alpha_q),\ v(\alpha_q) = v_q =
\sinh(\alpha_q),\\
u_q^2 - v_q^2 = 1.
\end{eqnarray}
This result is a simple consequence of the commutation relation
(\ref{rho_comm}). This identity can be derived by a variety of methods. For
example, one can differentiate its left-hand side with respect to $\alpha_q$ for
both $p={\rm L}$ and $p={\rm R}$ and solve the resultant differential
equation system.

The easiest way to transform the kinetic energy term is to
notice that the kinetic energy density can be expressed as a product of two
density operators. The derivation goes as follows. First, we normal order
the product of two density operators:
\begin{eqnarray}
\rho_p (x) \rho_p (y)&=&
\colon \psi_p^\dagger(x) \psi_p^{\vphantom{\dagger}}(x)
\psi_p^\dagger(y) \psi_p^{\vphantom{\dagger}}(y)\colon + 
s_p(x-y)\colon\psi_p^{\vphantom{\dagger}}(x) \psi_p^\dagger(y)\colon +
\label{rr}\\
&&s_p(x-y) \colon\psi_p^{\dagger}(x) \psi_p^{\vphantom{\dagger}}(y)\colon + 
b_p(x-y),
\nonumber \\
s_p(x)&=&\frac{p}{2\pi i\left(x-ip0\right)},\\
b_p(x)& = &\left(s_p(x)\right)^2,
\end{eqnarray}
As it is explained in Appendix A the normal ordering is used here to isolate
explicitly singular terms of the field operator products. Now we expand the
above identity into Laurent series with respect to powers of $(x-y)$:
\begin{eqnarray}
\rho_p (x) \rho_p (y)&=&b_p(x-y)
+\frac{ip}{2\pi} \colon\psi_p^{\vphantom{\dagger}}(x)
\nabla\psi_p^\dagger(x)\colon +
\frac{ip}{2\pi} \colon\psi_p^{\dagger}(x)
\nabla \psi_p^{\vphantom{\dagger}}(x)\colon 
+ (\text{irrelevant\ operators}). \label{Laurent}
\end{eqnarray}
In this expansion an irrelevant
operator can be recognized by a factor of $(x-y)^n$ where $n>0$. For
example, the first term in (\ref{rr}) is an irrelevant operator:
\begin{equation}
\colon\psi_p^\dagger(x) \psi_p^{\vphantom{\dagger}}(x)
\psi_p^\dagger(y) \psi_p^{\vphantom{\dagger}}(y)\colon \ \approx
\left(x-y\right)^2\colon\psi_p^\dagger(x) \psi_p^{\vphantom{\dagger}}(x)
\nabla \psi_p^\dagger(x) \nabla\psi_p^{\vphantom{\dagger}}(x)\colon.
\label{irrelevant}
\end{equation}
Indeed, by simple power counting one can verify that its scaling dimension is
equal to $d=4<2$. Sending $y \rightarrow x$ in eq. (\ref{Laurent}), we
establish the identity:
\begin{equation}
\frac{ip}{2\pi} \left( \colon\psi_p^\dagger(x)
(\nabla\psi_p^{\vphantom{\dagger}}(x))\colon  -
\colon(\nabla\psi_p^\dagger(x)) \psi_p^{\vphantom{\dagger}}(x)\colon\right)
=\lim_{y\rightarrow x} \left\{ \rho_p(x) \rho_p(y) -
b_p(x-y)\right\}.
\label{rhorho}
\end{equation}
Thus, it is permissible to write the hamiltonian (\ref{H}) in the form:
\begin{eqnarray}
H= \frac{\pi v_{\rm F}}{L} \sum_{pq} \left(\rho_{pq} \rho_{p-q} + 
\frac{g_q}{2\pi v_{\rm F}} \rho_{pq} \rho_{-p-q} \right), \label{Hr}
\end{eqnarray}
where $g_q$ is the Fourier transform of $\hat g(x)$. 
For convenience we explicitly show the zero mode part of the
above expression:
\begin{eqnarray}
H= \frac{\pi v_{\rm F}}{L} \sum_{q \ne 0} \sum_p \left(\rho_{pq} \rho_{p-q} + 
\frac{g_q}{2\pi v_{\rm F}} \rho_{pq} \rho_{-p-q} \right) +
\frac{ \pi v_{\rm F}}{L} \left( N_{\rm L}^2 + N_{\rm R}^2 \right) +
\frac{g_0}{L} N_{\rm L} N_{\rm R}.
\end{eqnarray}
Such splitting is useful since our transformation $U$ does not act on the
zero modes.
We now apply $U$ and choose parameters $\alpha_q$ in such a way that the
term $\rho_{\rm L} \rho_{\rm R}$ vanishes:
\begin{eqnarray} 
\tanh 2\alpha_q = -\frac{g_q}{2\pi v_{\rm F}},\\
u^2_q = \frac{1}{2} \left( 1 + \frac{1}{\sqrt{ 1 -
\left({g_q}/{2\pi v_{\rm F}}\right)^2}} \right),\quad
v^2_q = \frac{1}{2} \left( - 1 + \frac{1}{\sqrt{ 1 -
\left({g_q}/{2\pi v_{\rm F}}\right)^2}} \right), \quad v_q g_q < 0.
\label{uv}
\end{eqnarray}
The transformed hamiltonian is:
\begin{eqnarray}
U H U^\dagger&=&\frac{\pi v_{\rm F}}{L} \sum_{q \ne 0} \sum_p
\left(u_q^2 + v_q^2 +
\frac{g_q}{\pi v_{\rm F}} u_q v_q \right) \rho_{pq} \rho_{p-q} + 
\frac{ \pi v_{\rm F}}{L} \left( N_{\rm L}^2 + N_{\rm R}^2 \right) +
\frac{g_0}{L} N_{\rm L} N_{\rm R} \\
&=&\frac{\pi v_{\rm F}}{L} \sum_{pq} 
\left(u_q^2 + v_q^2 +
\frac{g_q}{\pi v_{\rm F}} u_q v_q \right) \rho_{pq} \rho_{p-q} + 
\frac{ \pi \left(v_{\rm F} - \tilde v_{\rm F} \right)}{L}
\left( N_{\rm L}^2 + N_{\rm R}^2 \right) +
\frac{g_0}{L} N_{\rm L} N_{\rm R}, \\
\tilde v_{\rm F}&=& v_{\rm F}
\sqrt{1-\left(\frac{g_0}{2\pi v_{\rm F}}\right)^2}.\label{tilde_v}
\end{eqnarray}
From now on we will ignore the zero mode contribution to the hamiltonian:
the above formulas clearly show that it is small ($\sim 1/L$) for a
macroscopic system. Ultimately, we have:
\begin{eqnarray}
U H U^\dagger&=&\frac{\pi v_{\rm F}}{L} \sum_{pq } 
\left(u_q^2 + v_q^2 +
\frac{g_q}{\pi v_{\rm F}} u_q v_q \right) \rho_{pq} \rho_{p-q}.
\label{Hfin}
\end{eqnarray}
This expression can be re-written in terms of fields $\psi$. Let us show how
this is done in a technically trivial case of
$\Lambda \rightarrow \infty$ (which is equivalent to $\hat g (x) =
g_0 \delta(x)$). When $\Lambda$ is infinite the functions $u_q$, $v_q$ and
$g_q$ are all constants independent of $q$. Their values are given by
(\ref{uv}) with $g_q \equiv g_0$. Therefore, hamiltonian (\ref{Hfin}) is
equal to
\begin{eqnarray}
U H U^\dagger&=&\frac{\pi \tilde v_{\rm F}}{L} \sum_{pq } 
\rho_{pq} \rho_{p-q},
\end{eqnarray}
where $\tilde v_{\rm F}$ is given by (\ref{tilde_v}). Finally, inverting
(\ref{rhorho}), we obtain
\begin{eqnarray}
U H U^\dagger = i \tilde v_{\rm F} \int dx \sum_p p \colon 
\psi_p^\dagger \nabla \psi_p^{\vphantom{\dagger}}\colon + {\rm const}.
\end{eqnarray}
These transformations, apart from minor differences, are equivalent to
the refermionization as it is done in \cite{vD-S}.

In certain situations, however, it is convenient to have finite $\Lambda$.
Then a subtler reasoning is required. It is possible to prove that:
\begin{eqnarray}
U H U^\dagger = i\tilde v_{\rm F} \int dx \sum_p p \colon \psi_p^\dagger
\nabla\psi_p^{\vphantom{\dagger}}\colon + ({\rm irrelevant\ operators}).
\label{Hqp}
\end{eqnarray}
Such proof is rather technical. It can be found in Appendix B.

Let us agree that operators with the tilde are transformations of
corresponding operators without the tilde, that is $\tilde O=
U^\dagger O U$. Then the hamiltonian can be expressed in terms of operators
$\tilde\psi$ as follows:
\begin{eqnarray}
H = i\tilde v_{\rm F} \int dx \sum_p p \colon \tilde \psi_p^\dagger
\nabla\tilde\psi_p^{\vphantom{\dagger}}\colon + ({\rm irrelevant\ operators}).
\label{H0}
\end{eqnarray}
The above equation is the desired mapping of TL model on the model of 
quasiparticles $\tilde\psi$ whose interactions are small irrelevant
operators. Due to the irrelevance of the interaction the low-energy
properties of $\tilde\psi$ are those of free fermions.

Using the developed framework it is possible to discuss the effect of
non-linear dispersion on the spectrum and correlations of (\ref{H}). We add
an extra term to the TL hamiltonian:
\begin{eqnarray}
H' = H + H_{\rm nl}, \\ 
H_{\rm nl} = \sum_p \int dx dx' \hat h(x-x') \nabla\psi^\dagger_p(x)
\nabla \psi_p^{\vphantom{\dagger}} (x'), \label{H2}
\end{eqnarray}
The subscript `nl' stands for `non-linear' dispersion. Function $h_q$
which is the Fourier transform of $\hat h(x)$ has these properties:
\begin{eqnarray}
h_{q} = v'_{\rm F} < v_{\rm F}/\Lambda\ {\rm for}\ |q|<\Lambda,
\label{q=0}\\
h_{q} < v_{\rm F}/|q| \ {\rm for}\ |q| > \Lambda.
\end{eqnarray}
The second of these two conditions guarantees that for
$v'_{\rm F} < v_{\rm F}/\Lambda$ the modified kinetic energy
$pv_{\rm F}q + h_q q^2$ has the same sign as the original kinetic energy 
$pv_{\rm F}q$. That is, $H_{\rm nl}$ does not induce an instability of the
ground state by creating spurious Fermi point.

Condition (\ref{q=0}) implies that for small $|q|<\Lambda$ it is permissible
to use
\begin{eqnarray}
H_{\rm nl} = v'_{\rm F} \sum_p \int dx \nabla \psi^\dagger_p (x) \nabla
\psi^{\vphantom{\dagger}}_p(x)
\end{eqnarray}
instead of (\ref{H2}). The quantity $2v'_{{\rm F}}$ is the Fermi velocity
derivative with respect to the momentum.

We need to express $H_{\rm nl}$ in terms of $\tilde\psi$. To do so it is
convenient to rewrite $H_{\rm nl}$ with the help of density operators $\rho_p$
rather than field operators $\psi_p$. This trick was already used in
(\ref{rhorho}). It is easy to establish that
\begin{eqnarray}
v'_{\rm F} \sum_p \colon (\nabla \psi^\dagger_p (x))
(\nabla \psi^{\vphantom{\dagger}}_p (x)) \colon - \frac{1}{6} \nabla^2
\rho_p(x) = \label{Hnl_aux}\\
\frac{2\pi v'_{\rm F}}{3} \sum_p \lim_{y \rightarrow x} \left\{ i p \rho_p (x)
\left[ \colon \psi^\dagger_p (y) (\nabla \psi^{\vphantom{\dagger}}_p (y))
\colon -
\colon (\nabla \psi^\dagger_p (y)) \psi^{\vphantom{\dagger}}_p (y) \colon
\right] - 4\pi b_p(x-y) \rho_p (y) \right\}. \nonumber
\end{eqnarray}
According to (\ref{rhorho}) the expression in the square brackets
proportional to the product of two density operators. Thus:
\begin{eqnarray}
{\cal H}_{\rm nl} (x) = 
v'_{\rm F} \sum_p \colon (\nabla \psi^\dagger_p (x))
(\nabla \psi^{\vphantom{\dagger}}_p (x)) \colon - \frac{1}{6} \nabla^2
\rho_p(x) = \label{rho3}\\
\frac{4\pi^2v'_{\rm F}}{3}
\sum_p \lim_{y \rightarrow x} \lim_{z \rightarrow y} \left\{
\rho_p (x) \left[ \rho_p (z) \rho_p(y) - b_p (z-y) \right] -
2b_p (x-y) \rho_p(y) \right\}. \nonumber
\end{eqnarray}
Now we have to substitute $\rho_{pq} = u \tilde \rho_{pq} + v \tilde
\rho_{-pq}$ for $q \ne 0$ in
the above expression. The resultant third order polynomial of $\tilde \rho$'s
must be rewritten in terms of $\tilde \psi$ with the help of (\ref{rr})
and (\ref{rho3}). The final expression for $H_{\rm nl}$ is:
\begin{eqnarray}
H_{\rm nl} = \sum_p \int dx \left\{
\tilde v'_{\rm F}\colon (\nabla \tilde \psi^\dagger_p)
(\nabla \tilde \psi^{\vphantom{\dagger}}_p ) \colon +
i p \tilde g' \tilde \rho_{-p} \left( \colon \tilde \psi^\dagger_p 
(\nabla \tilde \psi^{\vphantom{\dagger}}_p) \colon -
\colon (\nabla \tilde \psi^\dagger_p) \tilde \psi^{\vphantom{\dagger}}_p
\colon \right) + \tilde \mu \tilde \rho_p\right\}. \label{nlq}
\end{eqnarray}
The details of the derivation together with the exact formulas for
coefficients $\tilde v'_{\rm F}$, $\tilde g'$ and $\tilde \mu$ can be found
in Appendix C. Here we quote only the expressions which are valid if the
interaction parameter $g_0$ is small. Let us define $\alpha_0$ as
$\alpha_0 = \alpha_{q=0}$. When interaction is small
($\alpha_0 \approx g_0/4\pi v_{\rm F} \ll 1$) the coefficients are:
\begin{eqnarray}
\tilde v'_{\rm F} - v'_{\rm F} \approx v'_{\rm F} \alpha^2_0,\\
\tilde g' \approx 2\pi v'_{\rm F}\alpha_0
 = \frac{g_0 v'_{\rm F}}{2 v_{\rm F}},\\
\tilde \mu \approx \gamma \Lambda^2 v'_{\rm F} \alpha_0^2,
\end{eqnarray}
where $\gamma$ is a non-universal constant of order unity.

The total hamiltonian $H'$ is the sum of $H$, eq. (\ref{H0}), and
$H_{\rm nl}$.  Due to $H_{\rm nl}$ the quasiparticle dispersion becomes
non-linear (the first term of (\ref{nlq})). Also, $H_{\rm nl}$ introduces
additional interactions between the quasiparticles. The operators
corresponding to these interaction (the second term in (\ref{nlq})) are
${\cal O}(v'_{\rm F} g_0)$ and irrelevant. Therefore, they can be
neglected provided that 
\begin{eqnarray} 
\tilde g'\Lambda / v_{\rm F} \ll 1 \Leftrightarrow
v'_{{\rm F}} g_0 \Lambda \ll v_{\rm F}^2. \label{free}
\end{eqnarray}
The quasiparticles remain free even if the bare
dispersion is not linear as long as the bare interaction is sufficiently
small. Note, that the free boson representation for TL model is much
less tolerant of $H_{\rm nl}$. This term is equivalent to cubic interaction
between bosons (see (\ref{rho3}) and Ref.\cite{haldane}). For the bosons to
remain free a stricter condition $v'_{{\rm F}} \Lambda \ll v_{\rm F}$ has
to be satisfied.

Equations (\ref{H0}) and (\ref{nlq}) allow us to describe the thermodynamics
of the electronic liquid with the generic dispersion. For example, the free
energy equals to
\begin{equation}
F = T\sum_{pk} \log \left( 1 + {\rm e}^{-\tilde \varepsilon_p(k)/T} \right)
=\frac{LT}{\pi} \int \frac{d \tilde \varepsilon}
{\sqrt{ \tilde v^2 _{\rm F} + 4 \tilde v'_{\rm F} \tilde \varepsilon}}
\log \left( 1 + {\rm e}^{-\tilde \varepsilon/T} \right), \label{free_en}
\end{equation}
where $\tilde \varepsilon_p(k) = p \tilde v _{\rm F} k + \tilde v'_{\rm F} 
k^2$ is the dispersion of the quasiparticles. This result is
non-perturbative in $v'_{\rm F} \Lambda /v _{\rm F}$. 
Corrections to the free energy due to the neglected quasiparticle
interaction are ${\cal O} ((v'_{\rm F} g_0)^2)$. Thus, our result for $F$
is accurate provided that (\ref{free}) holds true. 

The specific heat at constant chemical potential can be derived from
(\ref{free_en}) in the limit of low temperature
($T \ll \tilde v_{\rm F}^2/ \tilde v'_{\rm F}$):
\begin{eqnarray}
C(T) = \frac{\pi}{3 \tilde v_{\rm F}} T + \frac{14 \pi^3
(\tilde v'_{\rm F})^2}{5 \tilde v_{\rm F}^5} T^3
+ {\cal O} \left(({\tilde v'_{\rm F} T}/{\tilde v_{\rm F}^2})^5 \right)
+  {\cal O} (g_0^4) + {\cal O} ((\tilde g')^2). \label{C(T)}
\end{eqnarray}
We observe that non-zero dispersion curvature generates $T^3$
contribution to the specific heat. (Note: irrelevant operators neglected
along the way also contribute to the specific heat. However, their
contribution is ${\cal O}(g_0^4)$ and ${\cal O} ((\tilde g')^2)$.
For small $g_0$ such corrections can be disregarded.)

The same result for the specific heat could be obtained with the help of
bosonization technique. There one must expand the free energy in orders of
$H_{\rm nl}$. These calculations are done in Appendix D. They provide an
important consistency check for the proposed approach.

\section{Non-rigorous overview of the method}

We have finished the introduction of our method and ready to apply it to
Green's function calculations. At this point we would like to make a break
between two rather technical parts and explain why the method works on
intuitive level. 

The central
issue of this discussion could be loosely formulated as follows: why
the strongly interacting bosons could be mapped on the weakly interacting
fermions, what kind of `cancellation' of interactions takes place? To answer
this question we direct our attention to most essential aspects of the
approach. All technical complications will be disregarded: we will ignore
zero modes, normal ordering and assume that $\Lambda = \infty$. This makes
the following presentation more transparent.

Let us start with the bosonized form of the TL hamiltonian, familiar to
many researchers \cite{haldane}:
\begin{eqnarray}
H'&=& H_{\rm kin} + H_{\rm int} + H_{\rm nl}, \label{Hbos}\\
H_{\rm kin} + H_{\rm int}&=& \int dx \frac{\tilde v_{\rm F}}{2} \left\{ 
{\cal K} \left( \nabla \Theta \right)^2 +
{\cal K}^{-1} \left( \nabla \Phi \right)^2 \right\}, \\
H_{\rm nl} &=& \int dx
\frac{2 \pi^2 v'_{\rm F}}{3\sqrt{2}}
\left\{ \left( \nabla \Theta + \nabla \Phi \right)^3
+ \left( \nabla \Theta - \nabla \Phi \right)^3 \right\}, \label{Hnl_bos}
\end{eqnarray}
where boson fields $\Theta$ and $\Phi$ are connected to the density
operators in the usual way:
\begin{eqnarray}
\rho_{p} = \frac{1}{\sqrt{2}} \left( \nabla \Phi + p \nabla \Theta \right).
\end{eqnarray}
Tomonaga-Luttinger liquid parameter ${\cal K}$ is defined below. For the
purpose of this Section it is enough to remember that for small $g_0$
parameter ${\cal K}$ is close to unity:
\begin{eqnarray}
1 - {\cal K} = {\cal O} (g_0).
\end{eqnarray}
As we see from (\ref{Hnl_bos}), the boson interaction is proportional to
$v'_{\rm F}$. Now we perform the Bogoliubov transformation:
\begin{eqnarray}
\tilde \Theta = {\cal K}^{1/2} \Theta,\\
\tilde \Phi = {\cal K}^{-1/2} \Phi.
\end{eqnarray}
Our hamiltonian becomes:
\begin{eqnarray}
H'&=&\int dx \frac{\tilde v_{\rm F}}{2} \left\{ 
\left( \nabla \tilde \Theta \right)^2 +
\left( \nabla \tilde \Phi \right)^2 \right\} \\
&+& \frac{2 \pi^2 v'_{\rm F}}{3\sqrt{2}}
\left\{ \left( {\cal K}^{-1/2} \nabla \tilde \Theta +
{\cal K}^{1/2} \nabla \tilde \Phi \right)^3 +
\left( {\cal K}^{-1/2} \nabla \tilde \Theta -
{\cal K}^{1/2}\nabla \tilde \Phi \right)^3 \right\}. \nonumber
\end{eqnarray}
We rewrite it as follows:
\begin{eqnarray}
H'&=&H_{\rm free} + \Delta H,\\
H_{\rm free}&=&\int dx \frac{\tilde v_{\rm F}}{2} \left\{ 
\left( \nabla \tilde \Theta \right)^2 +
\left( \nabla \tilde \Phi \right)^2 \right\} 
+ \frac{2 \pi^2 v'_{\rm F}}{3\sqrt{2}} \left\{ \left( \nabla \tilde \Theta +
\nabla \tilde \Phi \right)^3 +
\left( \nabla \tilde \Theta -
\nabla \tilde \Phi \right)^3 \right\} ,\label{Hfree}\\
\Delta H&=& \int dx \frac{2 \pi^2 v'_{\rm F}}{3\sqrt{2}} \left\{
\left( {\cal K}^{-1/2} \nabla \tilde \Theta + {\cal K}^{1/2}
\nabla \tilde \Phi \right)^3 +
\left( {\cal K}^{-1/2} \nabla \tilde \Theta - {\cal K}^{1/2}
\nabla \tilde \Phi \right)^3 \right.\\
&-& \left. \left( \nabla \tilde \Theta + \nabla \tilde \Phi \right)^3 -
\left( \nabla \tilde \Theta - \nabla \tilde \Phi \right)^3 \right\}.
\nonumber
\end{eqnarray}
We split the above hamiltonian into two part for a reason. The first part
$H_{\rm free}$ becomes a free quasiparticle hamiltonian upon
refermionization. Indeed, the first term in the curly brackets of
(\ref{Hfree}) 
refermionizes to become the kinetic energy of the fermions with linear
dispersion, the second term of (\ref{Hfree}) becomes $q^2$-correction
to the linear dispersion.

Refermionized hamiltonian $\Delta H$ contains all the quasiparticle
interaction. It also contains terms quadratic in $\tilde \psi$
which renormalize the value
of $v'_F$. The important point, however, is that $\Delta H$ is small if
$g_0$ is small:
\begin{eqnarray}
\Delta H = {\cal O} ((1 - {\cal K})v'_{\rm F}) = {\cal O} (g_0 v'_{\rm F})
\end{eqnarray}
This can be established by observing that $\Delta H$ vanishes when $g_0 =0$
$\Leftrightarrow$ ${\cal K} = 1$. Note also, that $\Delta H$ is small if
(\ref{free}) is satisfied.
Power counting shows that $\Delta H$ is irrelevant. Smallness and
irrelevance of the quasiparticle interaction implies that the
quasiparticles could be viewed as weakly interacting. 

We could look at our method from a different prospective. The hamiltonian
$H' = H_{\rm kin} + H_{\rm int} + H_{\rm nl}$ in addition to a quadratic
(in $\psi$)
part $H_{\rm kin}$ has small marginal operator $H_{\rm int}$. Plus, it has
an irrelevant operator $H_{\rm nl}$ which we want to account for
non-perturbatively. Since perturbation
theory in the marginal operator diverges, we must either sum certain
diagrams to all orders or, as it is done in the theory of
superconductivity, perform a Bogoliubov transformation which kills the
undesirable marginal operator. This is what our transformation $U$ does:
$U (H_{\rm kin} + H_{\rm int}) U^\dagger$ is quadratic in fermionic fields.

Operator $H_{\rm nl}$ is quadratic in $\psi$. In general, however, operator
$U H_{\rm nl} U^\dagger$ does not have to be quadratic. Thus, after the
removal of the marginal operator by the Bogoliubov transformation, new
interactions between the
quasiparticles are generated. Fortunately, they are irrelevant and small.
The smallness of the generated interactions becomes obvious if we observe
that for small $g_0$
transformation $U$ is close to the identity transformation: $U = 1 + {\cal
O}(g_0)$. Thus, the quasiparticle interaction induced by the action of
$U$ on $H_{\rm nl}$ could be treated perturbatively. This is the core of
the approach we proposed in the previous Section.

\section{Density-density propagator}
In this section we will calculate density-density Green's function.
The derivation of the correlation function for the total density operator
$\rho=\rho_{\rm L} + \rho_{\rm R}$ is quite simple if we note that $\rho$
is proportional to the quasiparticle density operator:
$\rho_q = (u_q + v_q)\tilde\rho_q$. Using this identity one obtains for
small $|q|$:
\begin{eqnarray}
{\cal D}_q (\tau) = - \frac{1}{L}
\left< T_\tau \left\{\rho_q(\tau) \rho_{-q}(0)\right\} \right> = 
{\cal K} \left(\tilde {\cal D}_{{\rm L}q}(\tau) +
\tilde {\cal D}_{{\rm R}q}(\tau)\right),\label{D}\\
{\cal K} = \left.\left(u_q + v_q\right)^2\right|_{q=0} = \sqrt{\frac{2\pi
v_{\rm F} - g_0}{2\pi v_{\rm F} + g_0}}.
\end{eqnarray}
The chiral Green's function $\tilde {\cal D}_{pq}(\tau) = -
\left< T_\tau \left\{ \tilde\rho_{pq}(\tau) \tilde \rho_{p-q}(0)
\right\}\right>/L$ will be calculated below to the zeroth order in the
quasiparticle interaction constant $\tilde g'$. Unlike expansion in orders
of $g_0$ the perturbative expansion in orders of $\tilde g'$ is a
well-defined procedure: as we explained, the quasiparticle-quasiparticle
interaction is irrelevant. With accuracy ${\cal O}((\tilde g')^0)$ the
propagator $\tilde {\cal D}_{pq}$ can be expressed as a convolution of two
single-quasiparticle propagators:
\begin{eqnarray}
\tilde {\cal D}_{pq}(\tau) = - \frac{1}{L}
\sum_k \left< T_\tau \left\{\tilde\psi^\dagger_{p(k+q)}
(\tau) \tilde\psi^{\vphantom{\dagger}}_{p(k+q)} (0) \right\} \right>
\left< T_\tau \left\{\tilde\psi^{\vphantom{\dagger}}_{pk} (\tau)
\tilde\psi^\dagger_{pk} (0) \right\} \right> ,\\
\tilde {\cal D}_{pq\omega} = \sum_k \frac{n(\tilde\varepsilon_p(k)) - 
n(\tilde\varepsilon_p(k+q))}{i\omega + \tilde
\varepsilon_p(k) - \tilde \varepsilon_p(k+q)}.\label{Dp}
\end{eqnarray}
In eq. (\ref{Dp}) $n(\varepsilon)$ is Fermi distribution function. Thus, at
$T=0$:
\begin{eqnarray}
\tilde {\cal D}_{pq\omega} = \frac{p}{4\pi \tilde v'_{\rm F}q} \log \left\{
\frac{i \omega - p \tilde v_{\rm F} q + \tilde v'_{\rm F} q^2}
{i \omega - p \tilde v_{\rm F} q - \tilde v'_{\rm F} q^2} \right\},\\
{\cal D}_{q\omega} =  \frac{{\cal K}} {4\pi \tilde v'_{\rm F} q}
\log \left\{ \frac{ \omega^2 + ( \tilde v_{\rm F} q -
\tilde v'_{\rm F} q^2)^2}{ \omega^2 + (\tilde v_{\rm F} q +
\tilde v'_{\rm F} q^2)^2 }\right\}.\label{Df}
\end{eqnarray}
This result has the accuracy of ${\cal O} ((v'_{\rm F} g_0)^2)$, as
explained above. The
omitted corrections are due to the quasiparticle interaction $\tilde g'$.
At the same time, the above expression for ${\cal D}$ is accurate to all
orders in the dispersion curvature $\tilde v'_{\rm F}$.

The retarded propagator $D_{q\omega}$ is obtained by the analytical
continuation:
\begin{eqnarray}
D_{q\omega} = \frac{{\cal K}} {4\pi \tilde v'_{\rm F} q}
\log \left\{ \frac{ ( \tilde v_{\rm F} q -
\tilde v'_{\rm F} q^2)^2 - (\omega + i0)^2}{ (\tilde v_{\rm F} q +
\tilde v'_{\rm F} q^2)^2 - (\omega +i0)^2}\right\}. \label{Dret}
\end{eqnarray}
For vanishing $\tilde v'_{\rm F}$ we can write for the Green's
function the following expansion:
\begin{eqnarray}
{ D}_{q\omega} =  \frac{\tilde v_{\rm F} {\cal K} q^2}{\pi
((\omega + i0)^2 - \tilde v_{\rm F}^2 q^2)} +
\frac{( \tilde v'_{\rm F})^2 {\cal K} q^5}{6\pi}
\left(\frac{1}{(\omega - \tilde v_{\rm F} q + i0)^3} -
\frac{1}{(\omega + \tilde v_{\rm F} q + i0)^3}\right) +
{\cal O} \left((v'_{\rm F})^4 \right) +
{\cal O} \left((v'_{\rm F}g_0)^2 \right).\label{D_exp}
\end{eqnarray}
The first term coincides with the well-known bosonization
result \cite{boson}. The second term could be also found within the
bosonization approach: one has to develop a perturbation theory expansion
in powers of $\tilde v'_{\rm F}$.
How this could be done is shown in Appendix E. These calculations serve as
yet another consistency check for our method.

One can make another interesting observation when examining (\ref{D_exp}).
By looking at this expansion it is impossible to guess that the propagator
(\ref{Dret})
has a branch-cut, not a pole. To determine the correct complex structure
the non-perturbative in $v'_{\rm F}$ calculations are required.

Such complex
structure of the propagator indicates that there is no coherently propagating
bosonic mode \cite{boson}. Instead, we have a continuum of quasiparticle -
quasihole pair excitations. It is possible to visualize this continuum 
by calculating the spectral function $B_{q\omega}$:
\begin{eqnarray}
B_{q\omega} = - 2 {\rm Im\/} D_{q\omega} =
\frac{{\cal K}} {2 \tilde v'_{\rm F} q} \left\{ 
\vartheta \left( \omega^2 - (\tilde v_{\rm F} q - \tilde v'_{\rm F} q^2)^2
\right) -
\vartheta \left( \omega^2 - (\tilde v_{\rm F} q + \tilde v'_{\rm F} q^2)^2
\right)
\right\} {\rm sgn\ } \omega.
\end{eqnarray}
(Note: unlike single-fermion spectral function $B_{q\omega}$ does not have to
be positive.) Every point of $(q,\omega)$ plane where $B_{q\omega} \ne 0$ 
corresponds to a quasiparticle - quasihole excitation with total energy
$|\omega|$ and total momentum $q$. The set of these points is shown on fig.2.

A previous attempt to account for non-zero $v'_{\rm F}$ has been made in
\cite{haldane,kopietz}. However, the bosonic representation used in the
latter references is not very convenient for such a task. To illustrate the
nature of the problem we consider a case of free electrons ($g_0=0$) with
strongly non-linear dispersion $v'_{\rm F} \Lambda \sim v_{\rm F}$:
\begin{eqnarray}
H_{\rm free} = H_{\rm kin} + H_{\rm nl}. \label{free_f}
\end{eqnarray}
The quasiparticle representation is trivial: $U = 1$ and $\psi = \tilde
\psi$. The spectrum of (\ref{free_f}) is 
\begin{eqnarray}
\varepsilon_k = v_{\rm F} |k| + v'_{\rm F} k^2. \label{E}
\end{eqnarray}
Density-density correlation function is given by (\ref{Df}). In the
bosonization framework operator 
$H_{\rm nl}$ corresponds to cubic interaction between bosons with the
dimensionless coupling constant $v'_{\rm F}\Lambda/v_{\rm F}$ of order unity
(see (\ref{Hnl_bos})). To compute either spectrum or Green's functions one
must resort to the perturbation theory whose accuracy, however, is not
obvious due to lack of small parameter.

If in addition to $v'_{\rm F}$ we have $g_0 \ne 0$ then the Bogoliubov
rotation of the TL bosons is required.
When the Bogoliubov transformation acts on $H_{\rm nl}$ it generates extra
interaction terms of the form $\tilde \rho_p^{\vphantom{2}}
\tilde \rho_{-p}^2$. The
coupling constant for this kind of interaction is of the order of
$v'_{\rm F} g_0$. Thus, in the bosonic representation one has to deal with
two kinds of interaction terms and two coupling constants one of which is
of the order of unity.

As we have seen in Section III, if we add weak
electron-electron interaction $H_{\rm int}$ (eq.(\ref{int})) to
$H_{\rm free}$ (eq.(\ref{free_f})) the quasiparticle representation
evolves continuously with $g_0$: $U=1 + {\cal O} (g_0)$,
$\tilde H' = H_{\rm free} + {\cal O}(g_0)$. Non-zero $g_0$ generates
interaction between the quasiparticles. Yet, this interaction remains
small (if (\ref{free}) is fulfilled) and irrelevant.
Consequently, the Green's function
is given by (\ref{Df}) with renormalized $v_{\rm F}$ and $v'_{\rm F}$, the
spectrum of the interacting hamiltonian has the form (\ref{E}). In
short, the advantage of the quasiparticle representation steams from
its ability to account for $v'_{\rm F}$ non-perturbatively.

\section{Single-electron Green's function}
Now we will show how to calculate the single-electron Green's function.
At present we can find the Green's function for the case of $v'_{\rm F} =
0$ only. Although, this Green's function has been evaluated before by a
number of approaches we would like to include this derivation to demonstrate
different aspects of the method.

In order to calculate the single-electron Green's function it is necessary
to know how to express the electron field operator $\psi^\dagger$ in terms
of the quasiparticle operators $\tilde\psi^\dagger$ and $\tilde\rho$. The
following derivation answers this question. We begin by introducing unitary
operator $\exp(\varphi_p(x))$ where
\begin{equation}
\varphi_p(x) = \sum_{q \ne 0} \frac{\beta_{pq}}{n_q} {\rm e}^{iqx}\rho_{pq},
\quad \beta_{pq} - {\rm real}\ {\rm coefficients}. 
\end{equation}
Let us calculate the commutator of this operator with $\rho_{pq}$:
\begin{equation}
\left[ {\rm e}^{\varphi_p(x)}, \rho_{pq} \right] =
\left( {\rm e}^{\varphi_p(x)} \rho_{pq} {\rm e}^{-\varphi_p(x)} -
\rho_{pq} \right) {\rm e}^{\varphi_p(x)}.
\end{equation}
The first term in brackets can be calculated easily:
\begin{equation}
{\rm e}^{\varphi_p(x)} \rho_{pq}{\rm e}^{-\varphi_p(x)} = \rho_{pq} + \left[
\varphi_p (x), \rho_{pq} \right] = \rho_{pq} + \beta_{p-q} p {\rm e}^{-iqx}.
\end{equation}
Thus, the formula for the commutator is:
\begin{equation}
\left[ {\rm e}^{\varphi_p}, \rho_{pq} \right] = p\beta_{p-q} {\rm e}^{-iqx}
{\rm e}^{\varphi_p}.
\end{equation}
If we choose $\beta_{pq}=p$ the product $\psi^\dagger_p {\rm e}^{\varphi_p}$
commute with $\rho_{p'q}$ for any $p'$ and $q \ne 0$. Therefore, it commutes
with $\Omega$, eq.(\ref{Omega}). This means that such product is invariant
under the action of $U$. The action of $U$ on the field operator is given by:
\begin{eqnarray}
\tilde \psi_p^\dagger = U^\dagger \psi_p^\dagger U= \psi_p^\dagger
{\rm e}^{\varphi_p} U^\dagger {\rm e}^{-\varphi_p} U.
\end{eqnarray}
The latter formula is easy to invert:
\begin{eqnarray}
\psi_p^\dagger(x) = \tilde \psi_p^\dagger(x) {\rm e}^{\tilde \varphi_p(x)}
U {\rm e}^{-\tilde \varphi_p(x)} U^\dagger
 = \tilde \psi_p^\dagger (x) \tilde {\cal F}_p^\dagger(x),\\
\tilde {\cal F}_p^\dagger(x) = \exp \left( -p \sum_{q \ne 0} \frac{1}{n_q}
{\rm e}^{iqx}
\left( w_q \tilde \rho_{pq} + v_q \tilde \rho_{-pq}
\right) \right),\\
\left[\tilde \psi_p^\dagger (x),
\tilde {\cal F}_p^\dagger (x) \right] = 0,\\
w_q = u_q - 1.
\end{eqnarray}
It gives us the desired equation for $\psi^\dagger$ in terms of $\tilde
\psi^\dagger$ and $\tilde \rho$.

Let us calculate the correlation function:
\begin{eqnarray}
\left< \psi_p^\dagger (x,\tau) \psi_p^{\vphantom{\dagger}} (0,0) \right> =
\left< \tilde\psi_p^\dagger (x,\tau) \tilde {\cal F}_p^\dagger (x,\tau)
\tilde {\cal F}_p^{\vphantom{\dagger}} (0,0)
\tilde \psi_p^{\vphantom{\dagger}} (0,0) \right>. \label{corr}
\end{eqnarray}
The simplest way to evaluate this expression is to transform it in the
following manner. The density operators with $pn_q<0 (pn_q>0)$ must be shifted
to the left (right) end, the quasiparticle field operators stay in the
middle. The reason
for such choice is that $\tilde \rho_{pq}\left|0\right>=0$ for $pn_q>0$ and 
$\left<0\right|\tilde \rho_{pq}=0$ for $pn_q<0$ where $\left| 0 \right>$ is
the ground state of (\ref{H}). The details of this
derivation are given in Appendix F. The result is
\begin{eqnarray}
\left< \psi_p^\dagger (x,\tau) \psi_p^{\vphantom{\dagger}} (0,0) \right>
=\left< \tilde \psi_p^\dagger (x,\tau)
\tilde \psi_p^{\vphantom{\dagger}} (0,0) \right> {\exp}\left\{  - \sum_q
(1-{\rm e}^{iqx- \tilde v_ {\rm F} |q|\tau})\frac{v_q^2}{\left|n_q\right|}
\right\}
= \frac{1}{2\pi(ipx- \tilde v_ {\rm F} \tau)}
\left( \frac{a^2}{x^2 + \tilde v_{\rm F}^2 \tau^2}\right)^\theta,\label{G}
\end{eqnarray}
where we used the notation $\theta=2v_q^2|_{q=0}^{\vphantom{2}} =
({\cal K} + 1/{\cal K} - 2)/2$.  The same formula is derived using
bosonization \cite{haldane,boson}.

It is well-known fact that the above Green's function does not have pole. This
means that the quasiparticle state has zero overlap with the physical electron
state.

\section{Discussion}

In this paper we solve the TL model with the help of the unitary
transformation. The transformation maps the original hamiltonian on the
hamiltonian of weakly interacting quasiparticles.

Our approach easily incorporates deviation of the electron
dispersion from the linear form. As long as the bare interaction constant is
sufficiently small two new non-perturbative results (eq. (\ref{free_en}) and
eq. (\ref{Df})) can be derived. The derived results accounts for the
curvature parameter $\tilde v'_{\rm F}$ to all orders; they contain
errors of order $(\tilde g')^2 \sim (v'_{\rm F} g_0)^2$ due to neglected
interaction among the quasiparticles.

The ability of our method to account for $\tilde v'_{\rm F}$ to all orders
allows us to resolve the complex structure of the density-density
propagator.

In principle, our diagonalization technique is not the only way to derive
the quasiparticle representation. It is possible (fig.1) to bosonize (\ref{H})
then perform the Bogoliubov transformation and then fermionize the diagonal
bosonic hamiltonian \cite{luther-emery,kivelson}. Alternatively,
one can construct non-fermionic excitations using appropriate exponents of
bosonic operators \cite{ng}. In our case use of bosonization looks like an
unnecessary detour.

To conclude, we propose a unitary transformation which maps TL model on
a model of free fermions. Such approach reproduces or generalizes the 
TL correlation functions calculated using bosonization.

\section{Acknowledgments}
The author is grateful to D.P. Arovas, A. Castro Neto, F. Guinea, A.J.
Millis for useful discussions. 

The author would like to thank the referees who
suggested writing Appendices D and E. 

Support of the ``Dynasty" foundation of
Dmitri Zimin is gratefully acknowledged.

\section{Appendix A}
In order to differentiate products of field operators one must perform 
normal ordering of the product under consideration. The subject of this
Appendix is the definition of the normal ordering procedure we use in this
paper.

Most commonly a normal ordered product of two or more field operators is
defined with the reference to some non-interacting ground state
$\left|0\right>$. For example, for product of two operators such definition
reads:
\begin{eqnarray}
\colon\psi^\dagger (x) \psi (x')\colon = \psi^\dagger (x) \psi (x') -
\left<0\right|\psi^\dagger (x) \psi (x') \left|0\right>. \label{Nord}
\end{eqnarray}
The ground state is not specified in this equation. Thus, we actually have
infinite number of normal ordering procedures, each corresponding to a
particular state $\left|0\right>$. Usually, the problem at hand dictates
the choice of this state.

Normal ordered product (\ref{Nord}) possesses two properties: (i) it is
well-defined and analytical at $x=x'$; (ii) its ground state expectation
value is zero.

Since the ground state of the Tomonaga-Luttinger hamiltonian cannot be
approximated by a ground state of non-interacting physical fermions $\psi$
TL ground state cannot be used in (\ref{Nord}). Consequently, it is
necessary to explain what state we use in our definition of the normal
order.

In the bosonization framework it is customary \cite{haldane,vD-S} to normal
order with respect to the the ground state of the hamiltonian:
\begin{eqnarray}
H_0 = \sum_k v_{\rm F} pkc^\dagger_k c^{\vphantom{\dagger}} _k. \label{Hk0}
\end{eqnarray}
We accept this definition in our work as well.
Thus, for a product of two field operators we have:
\begin{eqnarray}
\colon\psi^\dagger_p (x) \psi^{\vphantom{\dagger}}_p (x')\colon =
\psi^\dagger_p (x) \psi_p^{\vphantom{\dagger}}  (x') - s_p(x-x'),
\label{Nord'}\\
s_p(x) = \frac{p}{2\pi i (x -ip0)},
\end{eqnarray}
where $s_p$ is equal to $\left<0 \right| 
\psi^\dagger_p (x) \psi^{\vphantom{\dagger}}_p (x') \left|0 \right>$.
In this equation the pole of 
$\psi^\dagger_p (x) \psi^{\vphantom{\dagger}}_p (x')$ at $x=x'$ is
explicitly shown. The normal ordered product is well-behaved at $x=x'$ and
it satisfies the property (i), formulated above.

It is not a miracle that property (i) remains intact despite the fact that the
ground states of (\ref{H}) and (\ref{Hk0}) are drastically different. Note that
(i) is `ultraviolet' in its nature -- it refers to short distance
($|x-x'| \ll 1/\Lambda$), high energy ($|k| \gg \Lambda$) behavior
of the
operator product. Since interaction (\ref{int}) is limited in $q$-space
high-energy structure of the TL ground state is the same as that of
(\ref{Hk0}). This guarantees that (i) is satisfied.

Yet, the TL ground state expectation value of (\ref{Nord'}) for generic
values of $x$ and $x'$ is not zero. That is, property (ii) is violated.
Fortunately, we never need it in this paper.

The normal ordered product of four field operators is defined by the
equation:
\begin{eqnarray}
\psi_p^\dagger(x) \psi_p^{\vphantom{\dagger}}(x')
\psi_p^\dagger(y) \psi_p^{\vphantom{\dagger}}(y')&=&
\colon \psi_p^\dagger(x) \psi_p^{\vphantom{\dagger}}(x')
\psi_p^\dagger(y) \psi_p^{\vphantom{\dagger}}(y')\colon + 
s_p(x-y')\colon\psi_p^{\vphantom{\dagger}}(x') \psi_p^\dagger(y)\colon +\\
&&s_p(x'-y) \colon\psi_p^{\dagger}(x) \psi_p^{\vphantom{\dagger}}(y')\colon + 
s_p(x-y')s_p(x'-y).
\end{eqnarray}
As above, the normal ordered product can be differentiated everywhere.

For completeness, let us define the normal ordering of the density 
operators $\rho_p(x)$:
\begin{eqnarray}
\colon \rho_p (x) \rho_p (y) \colon&=&\rho_p (x) \rho_p (y) -
b_p (x-y),\label{Nrho2}\\
\colon \rho_p (x) \rho_p (y) \rho_p (z) \colon&=&\rho_p (x) \rho_p (y)
\rho_p (z) - b_p (x-y) \rho_p (z) - b_p (x-z) \rho_p (y) -\label{Nrho3}\\
&&b_p (y-z) \rho_p (x),\nonumber\\
b_p(x)&=&(s_p(x))^2.\label{bp}
\end{eqnarray}
We will use these definitions in Appendix C. 

The above rules can be
generalized for $T>0$: to normal order a product of density operators at
finite temperature one has to replace function $b_p(x)$ by function:
\begin{eqnarray}
b_{pT} (x)&=& \langle \tilde \rho_p(x) \tilde \rho_p(0) \rangle =
-\frac{T^2}{4 \tilde v_{\rm F}^2\sinh^2 (\pi T (x/ \tilde v_{\rm F} -ip0))}.
\label{bpT}
\end{eqnarray}
It is obvious that $b_{pT} = b_p$ at $T=0$.
In Appendix D we will need the finite $T$ definition.

\section{Appendix B}
In this Appendix we provide the derivation of (\ref{Hqp}) starting from the
formula (\ref{Hfin}). To express the
hamiltonian (\ref{Hfin}) in terms of $\psi$ rather than $\rho$ we introduce
a function $\Delta_q$:
\begin{eqnarray}
\Delta_q = u_q^2 + v_q^2 + \frac{g_q}{\pi v_{\rm F}} u_q v_q - 1 
\end{eqnarray}
and its Fourier transform $\hat \Delta(x)$:
\begin{eqnarray}
\hat \Delta(x) = \sum_q \left(u_q^2 + v_q^2 + \frac{g_q}{\pi v_{\rm F}}
u_q v_q - 1 \right) {\rm e}^{iqx}.
\end{eqnarray}
We defined $g_q$ such that $g_q \rightarrow 0$ when $|q|\rightarrow\infty$.
Thus, $v_q \rightarrow 0$, $u_q \rightarrow 1$ for $|q| \gg \Lambda$.
Therefore, the function $\Delta_q$
vanishes for large $|q|$. This implies that $\hat \Delta(x)$ vanishes for
$|x| \gg 1/\Lambda$. We may write:
\begin{eqnarray}
U H U^\dagger = 
\frac{\pi v_{\rm F}}{L} \sum_{pq} \rho_{pq} \rho_{p-q} +
\frac{\pi v_{\rm F}}{L} \sum_{pq} \Delta_q \rho_{pq} \rho_{p-q} =
\label{HDelta} \\
i v_{\rm F} \int dx \sum_p p :\psi_p^\dagger
\nabla\psi_p^{\vphantom{\dagger}}: +
\frac{\pi v_{\rm F}}{L} \sum_p
\int dx dx' \hat\Delta(x-x') \rho_p (x) \rho_p(x').
\nonumber
\end{eqnarray}
The first term in the above equation was obtained by inversion of
(\ref{rhorho}). The second term must be normal ordered first:
\begin{eqnarray}
\frac{\pi v_{\rm F}}{L} \sum_p \int dx dx' \hat\Delta(x-x')
\rho_p (x) \rho_p(x') =
\frac{\pi v_{\rm F}}{L} \sum_p \int dx dx' \hat\Delta(x-x') 
\left\{ b_p(x-x') + \colon \rho_p(x) \rho_p(x') \colon \right\}.
\end{eqnarray}
The additive constant proportional to $\int dx dx' \hat \Delta b_p$ will be
disregarded. The normal ordered product of
two $\rho$'s can be expanded into Taylor series in powers of $(x-x')$:
\begin{eqnarray}
\sum_p \int dx dx' \hat\Delta(x-x') \colon \rho_p (x) \rho_p(x')\colon &=&
\sum_p \left[ \int dx' \hat\Delta(x') \right] \int dx \colon
\rho_p^2 (x) \colon \label{aux}\\
&+&\frac{1}{2}\left[ \int dx' x'^2 \hat\Delta(x') \right]
\int dx \colon \rho_p (x) \nabla^2 \rho_p(x) \colon + \ldots,\nonumber
\end{eqnarray}
where ellipsis stand for higher order terms of the Taylor series. Since
$\hat \Delta (x)$ vanishes for large $|x|$ the integrals of $\hat \Delta$
are well-defined. We did not show the term proportional to
$\colon \rho_p \nabla \rho_p \colon$ since it is total derivative
$\nabla \colon \rho_p^2\colon$, thus, it vanishes upon integration.

We notice that the first term of expansion (\ref{aux}) is marginal. All
other terms
are irrelevant. Indeed, power counting shows that the scaling dimension of
the operator $\colon \rho_p \nabla^2 \rho_p \colon$ is $4 > 2$. The scaling
dimensions of the omitted terms are even higher. Thus, it is permissible
to write:
\begin{eqnarray}
\frac{\pi v_{\rm F}}{L} \sum_p \int dx dx' \hat\Delta(x-x')
\rho_p (x) \rho_p(x') = {\pi v_{\rm F}} \Delta_{0} \sum_p
\int dx \colon \rho_p^2(x) \colon + (\text{irrelevant\ operators}),\\
\Delta_0 = \Delta_q\big|_{q=0}.
\end{eqnarray}
Consequently, transforming $\colon \rho_p^2 \colon$ into the kinetic energy
operator with the help of (\ref{rhorho}) and substituting the result into
(\ref{HDelta}), we get the desired equation:
\begin{eqnarray}
U H U^\dagger & =&
\int dx \sum_p ip\tilde v_{\rm F} : \psi_p^\dagger
\nabla\psi_p^{\vphantom{\dagger}}: + ({\rm irrelevant\ operators}), \\
\tilde v_{\rm F}&=&v_{\rm F} ( 1 + \Delta_{0} ) = 
v_{\rm F}\left.\left(u_q^2 + v_q^2 + \frac{g_q}{\pi v_{\rm F}} u_q v_q
\right)\right|_{q=0} = v_{\rm F}
\sqrt{1-\left(\frac{g_0}{2\pi v_{\rm F}}\right)^2}.
\end{eqnarray} 
Note, that the omitted operators, in addition to being irrelevant, are also
small ($\sim \Delta_0 \sim g_0^2/v_{\rm F}^2$) if interaction is small
$g_0 \ll v_ {\rm F}$.

\section{Appendix C}

In this Appendix we show how $H_{\rm nl}$ can be expressed in terms of the
quasiparticle field operators $\tilde \psi^\dagger$, $\tilde \psi$.
We start with expression for the local density ${\cal H}_{\rm nl}$:
\begin{eqnarray}
{\cal H}_{\rm nl} (x)&=&v'_{\rm F} \sum_p \colon (\nabla \psi_p^\dagger(x))
(\nabla \psi_p(x)) \colon - \frac{1}{6}\nabla^2 \rho_p (x) = \label{Hnl1}\\
&&\frac{4\pi^2v'_{\rm F}}{3} \sum_p \lim_{z \rightarrow x}
\lim_{y \rightarrow x} \left\{
\rho_p (z) \left[ \rho_p (x) \rho_p(y) - b_p (x-y) \right] -
2b_p (z-y) \rho_p(y) \right\}. \nonumber
\end{eqnarray}
The last line can be abbreviated if one use the definition of the normal
ordering for the density operators $\rho_p (x)$, eq. (\ref{Nrho3}).
With this notation we can rewrite the formula for ${\cal H}_{\rm nl}$:
\begin{eqnarray}
{\cal H}_{\rm nl} = \frac{4\pi^2 v'_{\rm F}}{3} \sum_p \colon \rho_p^3 \colon.
\label{Hnl2}
\end{eqnarray}
To express the density operator $\rho_p (x)$ in terms of the quasiparticle
density operators $\tilde \rho_p (x)$ we must remember that:
\begin{eqnarray}
\rho_{pq} = u_q \tilde \rho_{pq} + v_q \tilde \rho_{-pq}\text{\ for\ }
q \ne 0,\\
N_p = \tilde N_p \text{\ for\ zero\ modes.}
\end{eqnarray}
Therefore, the density operator in co-ordinate space is equal to:
\begin{eqnarray}
\rho_p (x) = \delta \tilde \rho_{p0} +
 \tilde \rho_p^u (x) + \tilde \rho_{-p}^v (x),\\
\delta \tilde \rho_{p0} = \left.
\left( (1 - u_q) \tilde N_p - v_q \tilde N_{-p} \right)\right|_{q=0},
\end{eqnarray}
where $\tilde \rho_p^{u}$ is the convolution of the density operator
$\tilde \rho_p (x)$ with $\hat u(x)$:
\begin{eqnarray}
\tilde \rho_p^u (x) = (\hat u*\tilde \rho_p )(x) =
\int dx' \hat u(x-x') \tilde \rho_p (x'),\\
\hat u(x) = \int \frac{dq}{2\pi} u_q {\rm e}^{iqx}.
\end{eqnarray}
Likewise, $\tilde \rho_{-p}^v = (\hat v*\tilde \rho_{-p}) (x)$, where
function $\hat v(x)$ is defined in the same manner as $\hat u(x)$. It is
tempting to equate ${\cal H}_{\rm nl}$ and
\begin{eqnarray}
\frac{4\pi^2 v'_{\rm F}}{3} \sum_p \colon \left( \delta \tilde \rho_{p0} 
+ \tilde \rho_p^u + \tilde \rho_{-p}^v \right)^3 \colon.
\end{eqnarray}
This, however, is not exactly accurate since the normal ordering and
Bogoliubov transformation do not `commute' with each other. Such phenomena
becomes obvious if we were to think in terms of Bose creation and
annihilation operators. Normal ordering places all creation operators to the
left of all annihilation operators. Bogoliubov transformation mixes creation
and annihilation operators, thus, it spoils normal ordering. 

To rewrite ${\cal H}_{\rm nl}$ in terms of $\tilde \rho^{u,v}$ correctly let
us first examine the expression in square brackets in eq. (\ref{Hnl1}):
\begin{eqnarray}
\rho_p (x) \rho_p (y) - b_p (x-y) =
\colon \rho_p (x) \rho_p (y) \colon =
\left( \delta \tilde \rho_{p0} + \tilde \rho_p^u (x) + \tilde \rho_{-p}^v (x)
\right)  \left( \delta \tilde \rho_{p0} + \tilde \rho_p^u (y) +
\tilde \rho_{-p}^v (y) \right) - b_p (x-y) = \\
\left\{ \colon \tilde \rho_p^u (x) \tilde \rho_{p}^u (y) \colon +
(\hat u * b_p *\hat u) (x-y) - b_p (x-y) \right\} +
\left\{ \colon \tilde \rho_{-p}^v (x) \tilde \rho_{-p}^v (y) \colon +
(\hat v * b_{-p} * \hat v) (x-y) \right\} + \nonumber\\
\left\{\tilde \rho_p^u (x) \tilde \rho_{-p}^v (y) +
\tilde \rho_{-p}^v (x) \tilde \rho_{p}^u (y) \right\} +
\delta \tilde \rho_{p0} \left\{ \tilde \rho_p^u (x) + \tilde \rho_{-p}^v (x)
+ \tilde \rho_p^u (y) + \tilde \rho_{-p}^v (y) \right\} +
\delta \tilde \rho_{p0}^2 , \nonumber
\end{eqnarray}
where we applied the following transformations:
\begin{eqnarray}
\tilde \rho^u_p (x) \tilde \rho^u_p (y) = \int dx' dy' u(x-x') u(y-y') \tilde 
\rho_p (x') \tilde \rho_p (y')\\
 = \int dx' dy' u(x-x') u(y-y') \left(
\colon \tilde \rho_p (x') \tilde \rho_p (y') \colon + b_p (x'-y') \right)
\nonumber\\
= \colon \tilde \rho_p (x) \tilde \rho_p (y) \colon + (\hat u*b_p*\hat u).
\nonumber
\end{eqnarray}
Similar transformations could be done for $\tilde \rho^v(x) \tilde \rho^v(y)$.
Since $u_q^2 - v_q^2 = 1$ the identity:
\begin{eqnarray}
\hat u*f*\hat u = f + (\hat v*f*\hat v)
\end{eqnarray}
holds true. Therefore:
\begin{eqnarray}
\colon \rho_p (x) \rho_p (y) \colon = 
\colon \left( \tilde \rho_p^u (x) + \tilde \rho_{-p}^v (x) \right)
\left( \tilde \rho_p^u (y) + \tilde \rho_{-p}^v (y) \right) \colon
+ \left( \hat v * (b_p + b_{-p}) * \hat v \right)(x-y) + \\
\delta \tilde \rho_{p0} \left\{ \tilde \rho_p^u (x) + \tilde \rho_{-p}^v (x)
+ \tilde \rho_p^u (y) + \tilde \rho_{-p}^v (y) \right\} +
\delta \tilde \rho_{p0}^2 , \nonumber
\end{eqnarray}
The extra terms in this formula are non-singular functions of $(x-y)$. In
particular:
\begin{eqnarray}
\left( \hat v * (b_p + b_{-p}) * \hat v \right) (x-y)=
\int \frac{dq}{(2\pi)^2} {\rm e}^{iq(x-y)} |q| v_q^2.
\end{eqnarray}
Observe that the normal ordered product of $\rho$'s differs from the normal
ordered product of $\tilde \rho$'s by a non-singular operator.

Generalizing the above calculations for the product of three $\rho$'s 
we obtain the following expression for ${\cal H}_{\rm nl}$:
\begin{eqnarray}
{\cal H}_{\rm nl} = \frac{4\pi^2v'_{\rm F}}{3}
\sum_p \colon (\tilde \rho^u_p)^3
\colon + \colon (\tilde \rho^v_{-p})^3 \colon +
3 \tilde \rho^u_p \colon ( \tilde \rho_{-p}^v )^2 \colon +
3 \tilde \rho^v_{-p} \colon ( \tilde \rho_{p}^u )^2 \colon +
\tilde c\left(\delta \tilde \rho_{p0} + \tilde \rho^u_p + \tilde \rho_{-p}^v
\right) + (\text{z.m.}), \label{Hnl3}\\
\tilde c = \frac{3}{4\pi^2} \int dq |q| v_q^2,
\end{eqnarray}
where `z.m.' stand for zero modes terms which vanish in the
thermodynamic limit.

For small momenta the above formula can be simplified. What must be done,
in its substance, amounts to inversion of equations (\ref{Hnl1}) and
(\ref{rhorho}).
To illustrate this statement, consider the first and the second terms of 
(\ref{Hnl3}). Limiting ourselves to the case $\Lambda = \infty$ we can
neglect the $q$-dependence of $u_q$ and $v_q$. Then,
$\tilde \rho_p^u = u \tilde \rho_p$, $\tilde \rho_p^v = v \tilde \rho_p$
and it is possible to write:
\begin{eqnarray}
\frac{4\pi^2v'_{\rm F}}{3}
\sum_p \colon (\tilde \rho^u_p)^3 \colon + \colon (\tilde \rho^v_p)^3 \colon
= \frac{4\pi^2v'_{\rm F}}{3} (u^3 + v^3) \sum_p \colon \tilde \rho_p^3
\colon =  v_{\rm F}'  (u^3 + v^3)
\left( \colon \nabla \tilde \psi_p^\dagger \nabla \tilde
\psi_p^ {\vphantom{\dagger}} \colon - \frac{1}{6} \nabla^2 \tilde \rho_p
\right). \label{finite_Lambda}
\end{eqnarray}
When handling the third and the fourth terms of (\ref{Hnl3}) we are to act in
the similar fashion: the product $\colon \tilde \rho_p^2 \colon$ should be
rewritten in terms of $\tilde \psi$ with the help of (\ref{rhorho}). These
two terms give the quasiparticle interaction.

For a finite value of $\Lambda$ our task becomes somewhat more complicated.
In such a situation one can generalize the procedure of Appendix B.  Let us
briefly discuss the core of this generalization. As an example, consider
the expression:
\begin{eqnarray}
\colon (\tilde \rho_p^v)^3 \colon  = \int dx' dx'' dx''' v(x-x') v(x-x'')
v(x-x''')
\colon \tilde \rho_p (x')\tilde \rho_p (x'')\tilde \rho_p (x''') \colon.
\end{eqnarray}
It is easy to show with the help of Taylor expansion (see Appendix B) that
the normal ordered product 
$\colon \tilde \rho_p (x')\tilde \rho_p (x'')\tilde \rho_p (x''') \colon $
is equal
to $\colon \tilde \rho_p^3 (x)\colon + (\text{more\ irrelevant\ operators})$,
where `more irrelevant operators' stands for operators whose scaling
dimension is bigger than 3.
Therefore, we obtain the following equation:
\begin{eqnarray}
\colon (\tilde \rho_p^v)^3 \colon = a\colon \tilde \rho_p^3
\colon + (\text{more\ irrelevant\ operators}),
\end{eqnarray}
where coefficient $a$ is equal to $v_q^3 |_{q=0}$. The expression 
$\colon \tilde \rho_p^3 \colon$ has to be transformed further as in eq.
(\ref{finite_Lambda}). The remaining terms of (\ref{Hnl3})
are transformed similarly.
Therefore, one establishes:
\begin{eqnarray}
{\cal H}_{\rm nl}&=&
\sum_p \left\{ \tilde v'_{\rm F} \colon (\nabla \tilde \psi_p^\dagger)
(\nabla \tilde \psi_p) \colon - \frac{\tilde v'_{\rm F}}{6}\nabla^2 \tilde
\rho_p
+ i p \tilde g' \tilde \rho_{-p} \left( \colon \tilde \psi^\dagger_p 
(\nabla \tilde \psi^{\vphantom{\dagger}}_p) \colon -
\colon (\nabla \tilde \psi^\dagger_p) \tilde \psi^{\vphantom{\dagger}}_p
\colon \right)\right\} +  \label{Hnl} \\
&&\tilde \mu ( \tilde \rho_{\rm L} +
\tilde \rho_{\rm R} ) +
(\text{more\ irrelevant\ operators}), \nonumber\\
\tilde v'_{\rm F}&=&v'_{\rm F} \left. \left( u_q^3 + v_q^3 \right)
\right|_{q=0},\\
\tilde g'&=&2\pi v'_{\rm F} \left.\left( u_q^2 v_q^{\vphantom{2}} +
u_q^{\vphantom{2}} v_q^2 \right) \right|_{q=0},\\
\tilde \mu&=& v'_{\rm F} \left(\int dq |q| v_q^2 \right).
\end{eqnarray}
The values of $u_q$ and $v_q$ are given by (\ref{uv}). Finally, we find for
the hamiltonian:
\begin{eqnarray}
H_{\rm nl} = \sum_p \int dx \left\{
\tilde v'_{\rm F}\colon (\nabla \tilde \psi^\dagger_p)
(\nabla \tilde \psi^{\vphantom{\dagger}}_p ) \colon +
i p \tilde g' \tilde \rho_{-p} \left( \colon \tilde \psi^\dagger_p 
(\nabla \tilde \psi^{\vphantom{\dagger}}_p) \colon -
\colon (\nabla \tilde \psi^\dagger_p) \tilde \psi^{\vphantom{\dagger}}_p
\colon \right) + \tilde \mu \tilde \rho_p\right\} + (\text{more\
irrelevant\ operators}).
\end{eqnarray}
We omitted the full derivative $\nabla^2 \rho$ from the hamiltonian since it
contributes on the system boundary only.

We don't want to prove here that the quasiparticle interaction produces no
ultraviolet divergence of the perturbation theory. Instead, this issue is
examined at the end of Appendix F. This is because the formalism developed
there is particularly convenient for discussion of ultraviolet properties of
the quasiparticle interaction.

Finally, let us briefly explain, why the quasiparticle interaction
generated by the action of $U$ on $H_{\rm nl}$ is small. This explanation
is equivalent to the one given in Sec.III. Initially, our $H_{\rm
nl}$ could be expressed as a cube of $\rho_p(x)$, eq. (\ref{Hnl2}).
(Although, superficially, this expression appears to be sixth order in
$\psi_p$, a subtler analysis shows that it is quadratic in $\psi_p$, eq.
(\ref{Hnl1}).) The
Bogoliubov transformation $U$ converts (\ref{Hnl2}) into (\ref{Hnl3}).
Unlike (\ref{Hnl2}), eq. (\ref{Hnl3}) is indeed a sixth order polynomial in
$\psi_p$. However, for small $g_0$ operator $U$ is close to the identity
transformation: $U= 1 + {\cal O}(g_0)$. Therefore,
\begin{eqnarray}
v'_{\rm F} \colon \rho^3_p \colon = v'_{\rm F}\colon \tilde \rho^3_p \colon
+ {\cal O}(g_0v'_{\rm F}).
\end{eqnarray}
The first term on the right is proportional to $\nabla \tilde
\psi_p^\dagger \nabla \tilde \psi_p^{\vphantom{\dagger}}$, eq. (\ref{Hnl1}).
All generated terms, such as quasiparticle interactions and corrections to
$v'_{\rm F}$, are ${\cal O}(g_0v'_{\rm F})$.
                                                                                
\section{Appendix D}
In this Appendix we show how the result for the specific heat,
eq.(\ref{C(T)}), can be obtained within the bosonization framework. More
specifically, ${\cal O}((v'_{\rm F})^2)$ correction to the specific heat
will be calculated with the help of the perturbation theory for the free
boson hamiltonian:
\begin{eqnarray}
H = \frac{\pi \tilde v_{\rm F}}{2} \sum_{pq} \tilde \rho_{pq} \tilde
\rho_{p-q}.\label{Hbos0}
\end{eqnarray}
We will see that it is the same as 
${\cal O}((v'_{\rm F})^2)$ term in eq.(\ref{C(T)}). 

Our first step is to rewrite the non-linear dispersion hamiltonian $H_{\rm
nl}$ in a form particularly suitable for $T>0$ calculations: since in this
Appendix we work at finite temperature it is natural to express $H_{\rm nl}$
with the help of the finite temperature generalization of (\ref{Nrho2})
and (\ref{Nrho3}). As it was explained, such generalization is achieved by
placing $b_{pT}$, eq. (\ref{bpT}), instead of $b_p$, eq.(\ref{bp}).

Using the new rules of the normal ordering we could prove that at
$T\ge 0$ the following is true:
\begin{eqnarray}
\tilde v'_{\rm F} \sum_p \colon (\nabla \tilde 
\psi_p^\dagger(x)) (\nabla \tilde \psi_p(x)) \colon - \frac{1}{6}
\nabla^2 \tilde \rho_p (x) = 
\frac{4\pi^2 \tilde v'_{\rm F}}{3} \sum_p \colon \tilde \rho_p^3 \colon +
\frac{T^2}{4 \tilde v_{\rm F}^2}\tilde \rho_p,\label{HnlT_dens}
\end{eqnarray}
where $\colon\ldots\colon$ denotes here the finite temperature normal
order.
The proof of this equation is simple. Start from eq.(\ref{rho3}):
\begin{eqnarray}
\tilde v'_{\rm F} \sum_p \colon (\nabla \tilde 
\psi_p^\dagger(x)) (\nabla \tilde \psi_p(x)) \colon - \frac{1}{6}
\nabla^2 \tilde \rho_p (x) = \\
\frac{4\pi^2 \tilde v'_{\rm F}}{3}
\sum_p \lim_{y \rightarrow x} \lim_{z \rightarrow y} \left\{
\tilde \rho_p (x) \left[ \tilde \rho_p (z) \tilde \rho_p(y) -
b_p (z-y) \right] - 2b_p (x-y) \tilde \rho_p(y) \right\} = \nonumber\\
\frac{4\pi^2 \tilde v'_{\rm F}}{3}
\sum_p \lim_{y \rightarrow x} \lim_{z \rightarrow y} \left\{
\tilde \rho_p (x) \left[ \tilde \rho_p (z) \tilde \rho_p(y) - b_{pT} (z-y)
\right] - 2b_{pT} (x-y) \tilde \rho_p(y) \right\} + \nonumber\\
 3 \tilde \rho_p(x)
\lim_{y \rightarrow x} \left\{ b_{pT} (x-y) - b_p (x-y) \right\} .
\nonumber
\end{eqnarray}
The last limit is 
\begin{eqnarray}
\lim_{y \rightarrow x} \left\{ b_{pT} (x-y) - b_p (x-y) \right\} =
\frac{T^2}{12 \tilde v_{\rm F}^2}.
\end{eqnarray}
Thus, bosonized $H_{\rm nl}$ equals to
\begin{eqnarray}
H_{\rm nl} = \frac{4\pi^2 \tilde v'_{\rm F}}{3} \sum_p \int dx
\left( \colon \tilde \rho_p^3 \colon + \frac{T^2}{4 \tilde v_{\rm F}^2}
\tilde \rho_p \right) + {\cal O}(g_0 v'_{\rm F}). \label{HnlT}
\end{eqnarray}
This formula does not imply that $H_{\rm nl}$ depends on the temperature.
In this Appendix the symbol $\colon \ldots \colon$ denotes 
the finite $T$
generalization of the normal ordering, consequently, the expression
$\colon \tilde \rho^3 \colon$ varies with temperature. The $T^2$ term
compensates the temperature dependence of the normal ordered product
$\colon \tilde \rho^3 \colon$ so that $H_{\rm nl}$ is temperature
independent as it must be. We remind the reader that
such a way of writing is a matter of convenience.

In the above expression we ignore the quasiparticle interaction and the
correction to the chemical potential since we are interested in ${\cal
O}((v'_{\rm F})^2)$ corrections only while the neglected terms contribute
to the thermodynamics at ${\cal O} ((v'_{\rm F} g_0)^2)$.

The correction to the free energy due to $H_{\rm nl}$ is given by the usual
perturbation theory formula:
\begin{eqnarray}
\delta F = - \frac{1}{2} \int d\tau \langle H_{\rm nl} (\tau)
H_{\rm nl} (0) \rangle = - (\tilde v'_{\rm F})^{2} L \int d\tau dx \left(
\frac{16\pi^4}{9} \langle \colon \tilde \rho^3_{\rm L}(x,\tau) \colon
\colon \tilde \rho^3_{\rm L}(0,0) \colon \rangle + \frac{\pi^4 T^4}{9
\tilde v_{\rm F}^4} 
\langle \tilde \rho_{\rm L} (x,\tau) \tilde \rho_{\rm L}(0,0) \rangle \right).
\end{eqnarray}
Using Wick's theorem we find:
\begin{eqnarray}
\langle \colon \tilde \rho^3_{\rm L}(x,\tau) \colon
\colon \tilde \rho^3_{\rm L}(0,0) \colon \rangle  =
6 \langle \tilde \rho_{\rm L} (x,\tau) \tilde \rho_{\rm L} (0,0)
\rangle^3,\\
\langle \tilde \rho_{\rm L} (x,\tau) \tilde \rho_{\rm L} (0,0) \rangle = -
\frac{T^2}{4 \tilde  v_{\rm F}^2 \sinh^2(\pi T (x/\tilde v_{\rm F} - i\tau))}.
\end{eqnarray}
Therefore:
\begin{eqnarray}
\delta F = ( \tilde v'_{\rm F})^2 L \int d\tau dx \left(
\frac{\pi^4 T^6} {6 \tilde v_{\rm F}^6}
\frac{1}{\sinh^6 (\pi T (x/ \tilde v_{\rm F}
- i\tau))} + \frac{ \pi^4 T^6}{36 \tilde v_{\rm F}^6}
\frac{1}{\sinh^2 (\pi T (x/ \tilde v_{\rm F} - i\tau))} \right).
\end{eqnarray}
It is easy to calculate the integral 
\begin{eqnarray}
\int_{-\infty}^{+\infty}
\frac{dx}{ \sinh^2(\pi T (x/ \tilde v_{\rm F} - i\tau))} =
- \frac{2 \tilde v_{\rm F}}{\pi T}
\end{eqnarray}
for $0 < \tau < \beta$.
To find the integral of $1/\sinh^6$ we use a trick. One can check directly
that:
\begin{eqnarray}
\frac{1}{\sinh^{6} r} = \frac{8}{15}
\frac{1}{\sinh^2 r} + \frac{d^2 I(r)}{dr^2},\\
I(x) = \frac{1}{20}
\frac{1}{\sinh^4 r} -
\frac{2}{15} \frac{1}{\sinh^2 r}.
\end{eqnarray}
The integral of the full derivative is zero and we establish
for $0 < \tau < \beta$
\begin{eqnarray}
\int_{-\infty}^{+\infty}
\frac{dx}{\sinh^6 (\pi T (x/ \tilde v_{\rm F} - i\tau))} =
\frac{8}{15} \int_{-\infty}^{+\infty}
\frac{dx}{\sinh^2 (\pi T (x/ \tilde v_{\rm F} - i\tau))} =
- \frac{16 \tilde v_{\rm F}}{15 \pi T}.
\end{eqnarray}
Therefore, one finds the following corrections to the free energy, the 
entropy and the specific heat:
\begin{eqnarray} 
\delta F/L = - \frac{7\pi^3( \tilde v'_{\rm F})^2 T^4}
{30 \tilde v_{\rm F}^5},\\
\delta S/L = - \frac{\partial (\delta F/L)}{\partial T} = 
\frac{14\pi^3( \tilde v'_{\rm F})^2 T^3}{15 \tilde v_{\rm F}^5},\\
\delta C = T \frac{\partial (\delta S/L)}{\partial T} = 
\frac{14\pi^3( \tilde v'_{\rm F})^2 T^3}{5 \tilde v_{\rm F}^5}.
\end{eqnarray}
The last expression coincides with the second term of eq.(\ref{C(T)}) as it
should be.

\section{Appendix E}
In this Appendix we calculate ${\cal O}((v'_{\rm F})^2)$ correction
to the zero-temperature boson propagator. Such correction must coincide
with the second term of (\ref{D_exp}). 

The correction in question to the chiral Matsubara propagator comes from
$H_{\rm nl}$, eq.(\ref{HnlT}), with $T=0$. The appropriate Feynman
diagram is shown on fig.3. As one can see from eq.(\ref{D})
the corresponding expression is:
\begin{eqnarray}
\delta {\cal D}_p (x,\tau)= - \frac{{\cal K}}{2}
\left(\frac{4\pi^2 \tilde v'_{\rm F}}{3}\right)^2
\int d\tau' d\tau'' dx' dx'' \langle T_\tau \left\{ \tilde \rho_p (0,0)
\colon \tilde \rho_p^3 (x',\tau') \colon
\colon \tilde \rho_p^3 (x'',\tau'') \colon \tilde \rho_p(x,\tau) \right\}
\rangle_{\rm con- \atop nected} + {\cal O}((v'_{\rm F} g_0)^2).
\end{eqnarray}
The time-ordered average with respect to hamiltonian (\ref{Hbos0}) 
is equal to:
\begin{eqnarray}
\int d\tau' d\tau'' dx' dx'' 
\langle T_\tau \left\{ \tilde \rho_p (0,0)
\colon \tilde \rho_p^3 (x',\tau') \colon
\colon \tilde \rho_p^3 (x'',\tau'') \colon \tilde \rho_p(x,\tau) \right\}
\rangle_{\rm con- \atop nected} = \label{loop} \\
3\times 3 \times 2 \times 2 
\int d\tau' d\tau'' dx' dx'' \tilde {\cal D}_p^0(\tau', x')
\left( \tilde {\cal D}_p^0 (\tau'' - \tau', x'' - x') \right)^2 
\tilde {\cal D}_p^0 (\tau-\tau'',x-x'') ,\nonumber
\end{eqnarray}
where $\tilde {\cal D}_p^0(x,\tau)$ is the chiral free boson Green's
function
$
\tilde {\cal D}_p^0 (x,\tau) = - (1/L)
\langle T_\tau \left\{ \tilde \rho_p (0,0)
\tilde \rho_p (x,\tau) \right\}\rangle 
$ for hamiltonian (\ref{Hbos0}).

The numerical factor on the right-hand side of (\ref{loop})
corresponds to 36 possible ways
of contracting the density operators into a connected diagram presented on
fig.3. First, the external operator $\tilde \rho_p (0,0)$ could contract
with either of two vertexes; the operator $\tilde \rho_p(x,\tau)$ contracts
with the remaining vertex. This gives a factor of two. Second, a given
external
operator could contract with either of three density operators in a vertex.
This gives a factor of three for one external operator and another factor
of three for the second external operator. Finally, after contracting with
the external operators, every vertex has two uncontracted density
operators. For them there are two ways of contraction. Thus, we have yet
another factor of two.

In the Fourier space we can write:
\begin{eqnarray}
{\cal D}_{pq\omega} \approx {\cal K} \tilde {\cal D}_{pq\omega}^0 +
32 \pi^4 {\cal K} ( \tilde v'_{\rm F})^2
\left( \tilde {\cal D}_{pq\omega}^0 \right)^2 \tilde \Pi_{pq\omega},\\
\tilde {\cal D}_{pq\omega}^0 = \frac{1}{2\pi}
\frac{pq}{i\omega - p \tilde v_{\rm F} q},\\
\tilde \Pi_{pq\omega} = -T\sum_{\Omega} \int \frac{dk}{2\pi}
\tilde {\cal D}_{pk\Omega}^0 \tilde {\cal D}_{p(q-k)(\omega-\Omega)}^0
= \frac{1}{48 \pi^3} \frac{p q^3}{i\omega - p \tilde v_{\rm F} q}.
\end{eqnarray}
Therefore, for correction to the full propagator we write:
\begin{eqnarray}
\delta {\cal D}_{q\omega} = \sum_p \delta {\cal D}_{pq\omega} =
\frac{{\cal K}( \tilde v'_{\rm F})^2 q^5}{6\pi} \left(
\frac{1}{(i\omega - \tilde v_{\rm F} q)^3}
- \frac{1}{(i\omega + \tilde v_{\rm F} q)^3} \right).
\end{eqnarray}
After analytical continuation we recover the ${\cal O}((v'_{\rm F})^2)$
term of (\ref{D_exp}).

\section{Appendix F}
In this Appendix we provide a detailed derivation of the single-electron
Green's function for TL model, equation (\ref{G}). The formalism developed for
the Green's function is also a convenient tool to prove that the quasiparticle
perturbation theory has no ultraviolet divergences.

We start with equation (\ref{corr}). First, the product $\tilde
{\cal F}^\dagger_p \tilde {\cal F}^{\vphantom{\dagger}}_p$ has to be 
normal-ordered:
\begin{eqnarray}
\tilde {\cal F}^\dagger_p (x,\tau)
\tilde {\cal F}^{\vphantom{\dagger}}_p(0,0)&=&
{\rm e}^{{\bf W}_p (x + ipy,0)} {\rm e}^{-{\bf W}_p^\dagger (x + ipy,0)} 
{\rm e}^{{\bf V}_p (x - ipy,0)} {\rm e}^{-{\bf V}_p^\dagger (x - ipy,0)} \times
\label{FF} \\
&&{\rm e}^{-p\sum_q
\frac{1}{n_q} \left(1-{\rm e}^{iqx - pqy}\right) w_q^2 \vartheta(pq) }
{\rm e}^{p\sum_q \frac{1}{n_q} \left(1-{\rm e}^{iqx + pqy}\right)
v_q^2  \vartheta(-pq)}, \nonumber\\
&&{\bf W}_p(x,x') =
{p\sum_q \frac{1}{n_q} \left({\rm e}^{iqx'}-{\rm e}^{iqx  }\right)
w_q \tilde \rho_{pq} \vartheta(-pq)} ,\\
&&{\bf V}_{p}(x,x')=
{p\sum_q \frac{1}{n_q} \left({\rm e}^{iqx'}
- {\rm e}^{iqx }\right)
v_q \tilde \rho_{-pq}\vartheta(pq) }.
\end{eqnarray}
Here $y=\tilde v_{\rm F} \tau$.
This equation was derived with the help of commutation relations
(\ref{rho_comm}). In order to calculate $\tilde \rho(\tau)$ we used the
expression (\ref{Hr}).

Second, we transform the whole expression 
$\tilde \psi^\dagger_p (x,\tau) \tilde {\cal F}^\dagger_p (x,\tau)
 \tilde {\cal F}^{\vphantom{\dagger}}_p (0,0)
\tilde \psi^{\vphantom{\dagger}}_p(0,0)$. We shift $\exp({\bf W}_p)$
to the left past the
quasiparticle field operator $\tilde \psi^\dagger_p$. The exponent
$\exp(-{\bf W}^\dagger_p)$
is shifted to the very right. To perform these shifts the
commutation rule (\ref{psi_comm}) has to be used.
\begin{eqnarray}
&&\tilde \psi^\dagger_p (x,\tau) \tilde {\cal F}^\dagger_p (x,\tau)
 \tilde {\cal F}^{\vphantom{\dagger}}_p (0,0)
\tilde \psi^{\vphantom{\dagger}}_p(0,0) =  
\left\{{\rm e}^{{\bf W}_p(x + ipy,0)}
\left( 
\tilde \psi^\dagger_p(x,\tau) \tilde \psi^{\vphantom{\dagger}}_p (0,0)
\right) {\rm e}^{-{\bf W}_p^\dagger (x+ipy,0)}\right\}
\times \label{field}\\
&&\left({\rm e}^{{\bf V}_p(x - ipy,0)}
{\rm e}^{-{\bf V}_p^\dagger(x - ipy,0)} \right)
{\rm e}^{-p\sum_q \frac{1}{n_q} \left(1-{\rm e}^{iqx - pqy}\right)
\left(w_q^2 + 2w_q\right)\vartheta(pq)}
{\rm e}^{p\sum_q \frac{1}{n_q} \left(1-{\rm e}^{iqx + pqy}\right)
v_q^2  \vartheta(-pq)}. \nonumber
\end{eqnarray}
After taking the expectation value of this expression and recalling that
$w_q^2 + 2 w_q = v_q^2$ we obtain the desired expression for the correlation
function.

The above formula sets a convenient background for a rather general
discussion of the ultraviolet properties of the quasiparticle-quasiparticle
interactions. Imagine that we add the following term to the TL hamiltonian:
\begin{eqnarray}
H_{\rm extra} = \sum_p \int dx dx' \hat Z(x-x') \psi^\dagger_p (x)
\psi^{\vphantom{\dagger}}_p (x').
\end{eqnarray}
Operator $H_{\rm nl}$ is a particular case of $H_{\rm extra}$ with $\hat Z
= - \hat h''$.

With the help of (\ref{field}) it is easy to rewrite $H_{\rm extra}$ in terms
of the quasiparticle operators $\tilde \psi$ and $\tilde \rho$:
\begin{eqnarray}
H_{\rm extra} = \sum_p \int dx dx' \hat Z(x-x') \zeta(x-x') \left\{
{\rm e}^{{\bf W}_p(x,x')} \left( \tilde \psi^\dagger_p (x)
\tilde \psi^{\vphantom{\dagger}}_p (x') \right)
{\rm e}^{-{\bf W}_p^\dagger (x,x')} \right\}
\left( {\rm e}^{{\bf V}_{-p}(x,x')}
{\rm e}^{-{\bf V}_{-p}^\dagger(x,x')}\right),\label{Hextra}\\
\zeta(x) =
{\rm e}^{\sum_q \frac{1}{|n_q|} \left(1-{\rm e}^{iqx  }\right) v_q^2 }.
\end{eqnarray}
Function $\zeta(x)$ is well-defined for $|x| \ll 1/\Lambda$ and vanishes
algebraically for large $|x|$. Operator (\ref{Hextra}) is normal-ordered,
therefore, it is safe to expand the exponentials:
\begin{eqnarray}
H_{\rm extra} = \sum_{n, m, n', m'} H_{nmn'm'} =
 \sum_{n,m,n',m'} \int dx dx' \frac{\hat Z \zeta}{n!m!n'!m'!}
{\bf W}_p^n \tilde \psi^\dagger_p \tilde \psi^{\vphantom{\dagger}}_p
(-{\bf W}_p^\dagger)^m {\bf V}_p^{n'}(-{\bf V}_p^\dagger)^{m'}
\end{eqnarray}
Since ${\bf W} = {\cal O}(g_0^2)$ and ${\bf V} = {\cal O}(g_0)$ such
expansion is controlled by the smallness of $g_0$. 
We want to argue that any interaction term $H_{nmn'm'}$,
$n + m + n' + m' > 0$ does not lead to ultraviolet divergence of perturbation
theory. Consider, for example, the second order correction to the ground
state energy:
\begin{eqnarray}
\delta E_{nn'} =  - \sum_N \sum_{p_1\ldots p_{N}  \atop k_1\ldots k_{N}  }
\frac{ \left| \left< {p_1 \ldots p_N \atop k_1 \ldots k_N} \Big| H_{n0n'0}
\Big|0\right> \right|^2}{\sum_i \tilde \varepsilon (p_i) + 
\tilde \varepsilon (k_i)} \delta(\sum_i p_i + k_i). \label{2-corr}
\end{eqnarray}
Summation in the above formula goes over $N\leq n+n'+1$ quasiparticle
momenta $p_i$ and over $N$ quasihole momenta $k_i$. We include $m=m'=0$
terms only because other terms annihilate the ground state. Thus,
$m,m'\ne 0$ terms do not contribute to the second order ground state
correction. They do, however, contribute to higher order corrections and to
corrections to different propagators.

If the matrix element does not vanish at large momenta the expression
(\ref{2-corr}) suffers from the ultraviolet divergence. We will show that the
matrix element goes to zero if at least one $|p_i|$ or $|k_i|$ exceeds
$2(n+n')\Lambda$.

Let us examine definitions of ${\bf W}$ and ${\bf V}$. Since $v_q$ and $w_q$
vanishes for $|q| > \Lambda$ every individual operator ${\bf W,V}$ changes the
total momentum by no more than $\Lambda$. Thus, momentum change produced by
$({\bf W}^n {\bf V}^{n'})$ is no more than $(n+n') \Lambda$. We will use this
information in our next step.

First step in evaluating the matrix element from (\ref{2-corr}) is creation
of a quasiparticle-quasihole pair by applying $\tilde\psi^\dagger \tilde\psi$
product to $\left| 0 \right>$. Since the total momentum of all excitations is
zero the momentum of this pair cancels the momentum of
$({\bf W}^n {\bf V}^{n'})$. Thus, the pair momentum magnitude is limited by
$(n+n')\Lambda$. Two elementary excitations, quasiparticle and quasihole, of
which the pair is composed have their momenta bound by $(n + n')\Lambda$.

As we explained above, by acting on the ground state the product
$\tilde \psi^\dagger \tilde \psi$ creates an excited state with a
quasiparticle and a quasihole. On this state the operator 
$({\bf W}^n {\bf V}^{n'})$ acts creating yet another excited state. Every
individual operator ${\bf V}$ or ${\bf W}$ when acting on a state with finite
number of the elementary excitations can either (i) create another
quasiparticle-quasihole pair or (ii) replace an elementary excitation with
another one of the same chirality and charge but different momentum.

In case (i) both newly created elementary excitations have their momenta
bound by $\Lambda$. This is because single ${\bf W}$ or ${\bf V}$ operator
cannot make quasiparticle-quasihole excitation with total momentum higher than
$\Lambda$.

In case (ii) an individual operator ${\bf W}$ or ${\bf V}$ can change
momentum of an elementary excitation by no more that $\Lambda$. As above, this
is a consequence of $v_q$ and $w_q$ vanishing at $|q| > \Lambda$. Therefore,
the product $({\bf W}^n {\bf V}^{n'})$ cannot change momentum of an elementary
excitation by more than $(n+n')\Lambda$. Consequently, momentum of an
elementary excitation is bound by $2(n+n')\Lambda$.

This means that in (\ref{2-corr}) summation is effectively performed over a
sphere $|k_i| < 2 (n + n') \Lambda$, $|p_i| < 2 (n + n') \Lambda$. Thus, this
expression has no ultraviolet divergence.

The above argument can be easily generalized to other perturbation theory
formulas to prove that the perturbation theory for the quasiparticles has no
ultraviolet divergences.

The absence of the ultraviolet divergences, together with irrelevance of the
quasiparticle - quasiparticle interaction, implies that the quasiparticle
perturbation theory is well-defined.

\newpage

\begin{figure} [!t]
\centering
\leavevmode
\epsfysize=6cm
\epsfbox{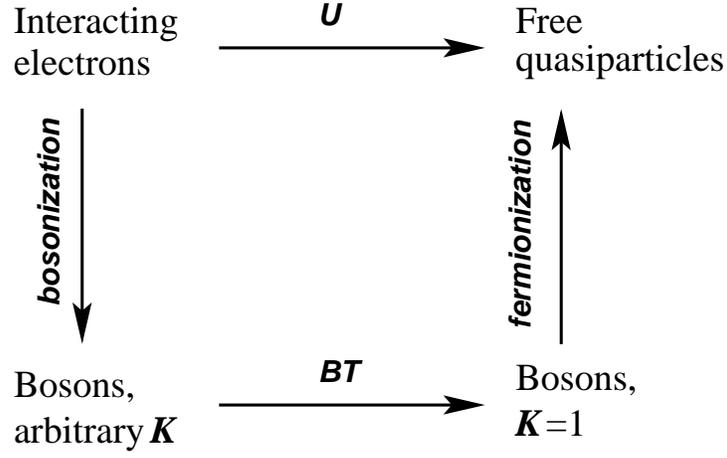}
\caption[]
{\label{fig1} 
Commutative diagram explains the relation between transformation $U$,
and the bosonization. `BT' stands for the Bogoliubov transformation of
bosons.
}
\end{figure}
\begin{figure} [!t]
\centering
\leavevmode
\epsfysize=5cm
\epsfbox{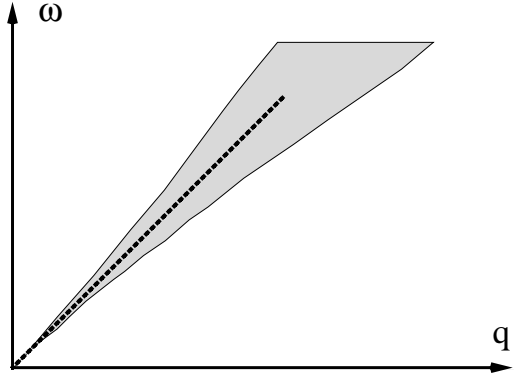}
\caption[]
{\label{fig2} 
When $v'_{\rm F} = 0$ the spectral density of $D_{q\omega}$ is delta-function
centered at $\omega = \tilde v_{\rm F} q$ line (dash line on the figure).
This is the dispersion curve of the TL bosons. For $v'_{\rm F} \ne 0$ the
spectral
density is non-zero within the whole shaded area. This area represents the
continuum of the quasiparticle-quasihole excitations. The TL bosons acquire
finite life-time in such a situation.
}
\end{figure}
\begin{figure} [!b]
\centering
\leavevmode
\epsfysize=4cm
\epsfbox{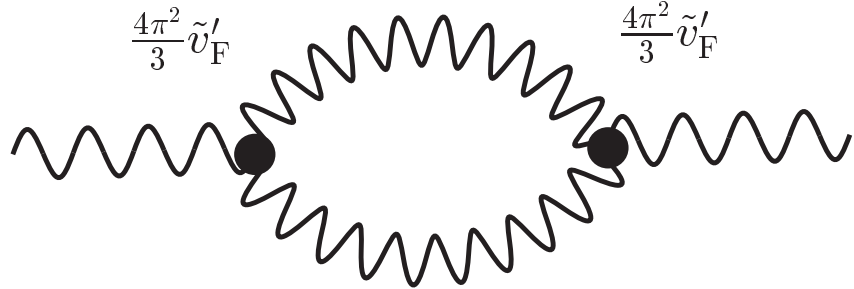}
\caption[]
{\label{fig3} 
Correction to $\tilde {\cal D}^0_{pq}$ due to the
dispersion curvature. Wavy lines correspond to $\tilde {\cal D}^0$, the
vertexes are proportional to $\tilde v'_{\rm F}$ \cite{jaxodraw}.
}
\end{figure}

\end{document}